\documentclass[prd, preprint, nofootinbib, showpacs]{revtex4}

\usepackage{epsf,epsfig,subfigure,graphicx,amsmath,amssymb}
\usepackage{amsfonts}
\usepackage{latexsym}
\usepackage{color}

\newcommand{\dis}[1]{\begin{equation}\begin{split}#1\end{split}\end{equation}}
\newcommand{\be}{\begin{equation}}
\newcommand{\ee}{\end{equation}}
\def\bea{\begin{eqnarray}}
\def\eea{\end{eqnarray}}

\newcommand{\eq}[1]{Eq.~(\ref{#1})}

\newcommand{\bfrac}[2]{{\left(\frac{#1}{#2} \right)  }}\newcommand{\VEV}[1]{\langle #1 \rangle}
\newcommand\gev{\,{\rm GeV}}
\newcommand\mev{\,{\rm MeV}}

\newcommand\mdm{{M_{\rm DM}}}

\begin{document}

\title{
 Light Dirac right-handed sneutrino dark matter}

\author{Ki-Young Choi}
%\email{kiyoung.choi@apctp.org}
 \affiliation{
 Asia Pacific Center for Theoretical Physics, Pohang, Gyeongbuk 790-784, Republic of Korea and \\
 Department of Physics, POSTECH, Pohang, Gyeongbuk 790-784, Republic of Korea}

\author{Osamu Seto}
% \email{seto@physics.umn.edu}
 \affiliation{
 Department of Life Science and Technology,
 Hokkai-Gakuen University,
 Sapporo 062-8605, Japan
}

%\date{\today}
%

%%%%%%%%%%%%%%%%%%%%%%
\begin{abstract}
%%%%%%%%%%%%%%%%%%%%%%
%
We show that mostly right-handed Dirac sneutrinos are a viable
 supersymmetric light dark matter candidate.
While the Dirac sneutrino scatters with nuclei dominantly
 through the $Z$-boson exchange and is stringently constrained
 by the invisible decay width of the $Z$ boson,
 it is possible to realize a large enough cross section with the nucleon to account for 
 possible signals observed at direct dark matter searches, such as CDMS II(Si) or CoGeNT. 
Even if the XENON100 limit is taken into account,
 a small part of the signal region for CDMS II(Si) events
 remains outside the region excluded by XENON100.
%
%%%%%%%%%%%%%%%%%%%%%%
\end{abstract}
%%%%%%%%%%%%%%%%%%%%%%

\pacs{}

\preprint{APCTP-Pre2013-006, HGU-CAP-023} 

\vspace*{3cm}
\maketitle

%==================================%
%          Main body               %
%==================================%

\section{Introduction}
\label{introduction}

Light weakly interacting massive particles (WIMPs)
 with masses around 10 GeV have received a lot of attention,
 motivated by the results
 of some direct dark matter (DM) detection experiments.
DAMA/LIBRA has claimed detection of the annual modulation
 signal by WIMPs~\cite{DAMALIBRA}. 
CoGeNT has found an irreducible excess~\cite{CoGeNT}
 and annual modulation~\cite{CoGeNTan}. 
CRESST has observed more events than expected backgrounds
 can account for~\cite{CRESSTII,Brown:2011dp}.
The CDMS II Collaboration has just announced~\cite{CDMSIISi} that
 their silicon detectors have detected three events and its possible signal region overlaps
 with the possible CoGeNT signal region analyzed by Kelso {\it et al.}~\cite{Kelso:2011gd}.
However, these observations are challenged by the null results
 obtained by other experimental collaborations,
 such as CDMS II~\cite{CDMSII, CDMSIIGe}, XENON10~\cite{XENON10},
 XENON100~\cite{XENON100:2011,XENON100:2012}
 and SIMPLE~\cite{SIMPLE}.
Recently, Frandsen {\it et al.}~\cite{Frandsen:2013cna} have pointed out that
 the XENON10 exclusion limit in Ref.~\cite{XENON10} 
 might be overconstraining.
It has been stressed that the signal region due to
 low-energy signals in CDMS II(Si)
 extends outside the XENON exclusion limit~\cite{DelNobile:2013cta}.

The Fermi-LAT collaboration has derived
 stringent constraints on the $s$-wave annihilation cross section
 of WIMPs by analyzing the gamma-ray flux from dwarf satellite galaxies~\cite{dSph}.
In particular, in the light-mass region below $ {\cal O} (10)$ GeV,
 the annihilation cross section times relative velocity $\langle\sigma v\rangle $
 of $ {\cal O}(10^{-26}) {\rm cm}^3/{\rm s}$, which corresponds to
 the correct thermal relic abundance $\Omega h^2\simeq 0.1$,
 has been excluded. 

Light WIMPs have been investigated as a dark matter interpretation
 of this positive data. 
In fact, very light neutralinos in the minimal supersymmetric Standard Model
 (MSSM)~\cite{Hooper:2002nq,Bottino:2002ry}
 and the next-to-MSSM (NMSSM)~\cite{Cerdeno:2004xw,Gunion:2005rw}
 or very light right-handed (RH) sneutrinos
 in the NMSSM~\cite{Cerdeno:2008ep,Cerdeno:2011qv,Choi:2012ba}
 have been regarded as such candidates.
However, these candidates hardly avoid the above Fermi-LAT
 constraint.~\footnote{If we give up the standard thermal relic, 
 we may consider a WIMP with an annihilation cross section small enough
 to satisfy the Fermi-LAT bound~\cite{Allahverdi:2013tca} or
 the nonvanishing dark matter asymmetry~\cite{Kang:2011wb,Okada:2013cba}.}

In this paper, we show that
 mostly right-handed Dirac sneutrinos are viable supersymmetric light
 DM candidates and
 have a large enough cross section with nucleons to account for 
 possible signals observed at direct DM searches.
Dirac sneutrinos scatter off nuclei dominantly
 via the $Z$-boson exchange process through the suppressed coupling
 and mostly with neutrons rather than protons. 
Although this $Z$-boson-mediated scattering does not relax the tension 
 among direct DM search experiments and
 its availability is limited by the invisible decay width of the $Z$ boson,
 a part of the signal region for CDMS II(Si) events~\cite{CDMSIISi}
 remains outside the excluded region by XENON100~\cite{XENON100:2012}.
We examine the cosmic dark matter abundance as well as the constraints from
 indirect dark matter searches 
 for a viable model of Dirac sneutrino dark matter.

The paper is organized as follows.
In Sec.~\ref{sneutrinoDM},
 we estimate the DM-nucleon scattering cross section through
 the $Z$-boson exchange process
 and show the experimental bounds and signal regions for this case.
We impose the bound from the $Z$ boson invisible decay width too.
In Sec.~\ref{other},
 after a brief description of the model, 
 we examine other cosmological, astrophysical, and
 phenomenological constraints.
We then summarize our results in Sec.~\ref{conclusion}.

\section{Dirac sneutrino dark matter direct detection}
\label{sneutrinoDM}

\subsection{Invisible $Z$-boson decay}

We are going to consider light Dirac sneutrino DM scattering with nuclei
 through the $Z$-boson exchange process in the direct detection experiments.
Since the property of the $Z$ boson is well understood,
 the possibility of a light sneutrino has been stringently constrained from 
 the invisible decay width of the $Z$ boson.
First, we briefly summarize the bound.
 
The $Z$-boson invisible decay is $(20.00\pm0.06) \%$ for the total decay width of
 the $Z$-boson decay $\Gamma_Z=2.4952\pm0.0023\gev$~\cite{PDG}. 
This gives a constraint on the neutrino number which couples to the $Z$ boson, given by~\cite{PDG}
\dis{
N_\nu =2.984\pm0.008, \qquad(\rm PDG).
}
% from Standard Model firs to LEP data~\cite{ALEPH:2005ab}.
%
% When we consider the corresponding invisible decay width is
%\dis{
%\Gamma_{\rm inv} \simeq 1.332\gev.
%}
%Or for one flavor neutrino
%\dis{
%\Gamma_{\rm inv, one flavor} \simeq 0.666\gev.
%}
The LEP bound on the extra invisible decay width is given as~\cite{ALEPH:2005ab}
\begin{equation}
 \Delta \Gamma_{\rm inv}^Z< 2.0 \mev \qquad (95\%\, \rm C.L.).
\label{ZinvBound}
\end{equation}

\begin{figure}[!t]
  \begin{center}
  \begin{tabular}{c}
   \includegraphics[width=0.6\textwidth]{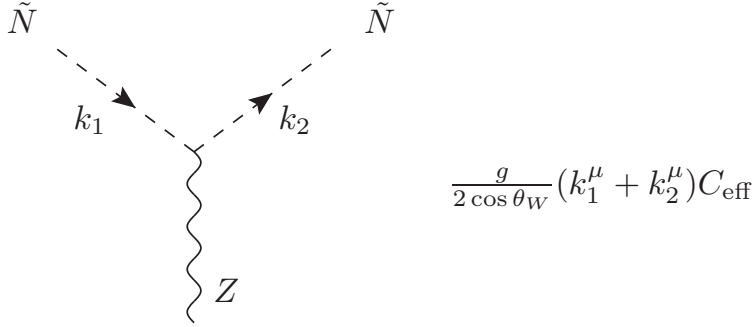}
  \end{tabular}
  \end{center}
  \caption{The effective vertex between a sneutrino and the $Z$ boson.}
  \label{fig:Zvertex}
\end{figure}

If there is a light sneutrino which couples to the $Z$ boson, the $Z$ boson can decay into light sneutrinos.
The spin-averaged amplitude is
\dis{
\overline{|M|^2} = \frac{|C_{\rm eff}|^2 g^2 M_Z^2}{12 \cos^2\theta_W}\left( 1-4\frac{M_{\tilde N}^2}{M_Z^2} \right) .
}
Here, $C_{\rm eff}$ parametrizes the suppression in the sneutrino-sneutrino-$Z$ boson coupling as shown in Fig.~\ref{fig:Zvertex}.
For pure left-handed sneutrinos, $C_{\rm eff}=1$. 
The decay width of the $Z$ boson into light sneutrino DM is given by
\dis{
\Gamma_{Z\rightarrow \tilde{N} \tilde{N}^*} &= \frac{|C_{\rm eff}|^2g^2 M_Z}{192 \pi \cos^2\theta_W}\left(1- 4\frac{M_{\tilde N}^2}{M_Z^2} \right)^{3/2},
}
 and we impose the upper bound (\ref{ZinvBound}) on this.
 This bound corresponds to
 \dis{
 C_{\rm eff} \lesssim 0.15,
 }
 for a few $\gev$ dark matter particle.
 The contour plot of the invisible decay width is also shown in Fig.~\ref{fig:ZinvAndDD}.

\subsection{Direct detection}

\begin{figure}[!t]
  \begin{center}
  \begin{tabular}{ccc c ccc}
   \includegraphics[width=0.3\textwidth]{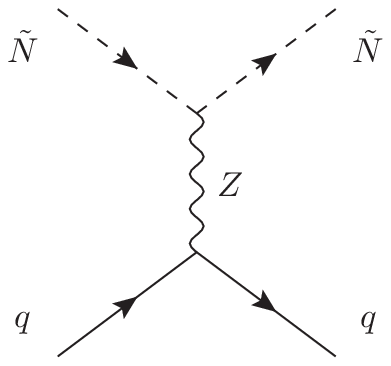}
    & & & &&&
  \includegraphics[width=0.3\textwidth]{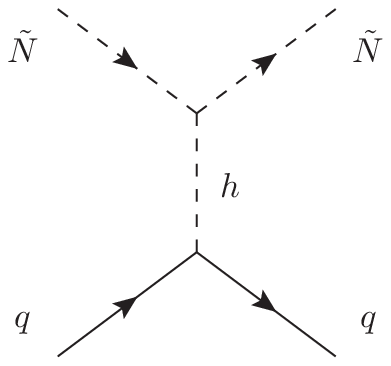}
  \end{tabular}
  \end{center}
  \caption{The diagrams for the elastic scattering of right-handed sneutrino dark matter with quarks.}
  \label{fig:DD}
\end{figure}

Dirac sneutrino DM can have elastic scattering with nuclei in the direct detection experiments.
The most relevant process is due to the $Z$-boson exchange as in the left diagram in Fig.~\ref{fig:DD}.
The $Z$-boson exchange cross section with nuclei ${}^A_ZN$ is given by
\begin{eqnarray}
\sigma^Z_{\chi N}&=&|C_{\rm eff}|^2 \frac{G_F^2}{2\pi} \frac{\mdm^2 m_N^2}{(\mdm+m_N)^2}
 \left[ A_N +2(2\sin^2\theta_W-1)Z_N \right]^2 \label{DDZ} \\
 &\simeq&   (A_N-Z_N)^2\bfrac{\mu_N^2}{\mu_n^2} \sigma^Z_{\chi n},
 \label{DD_Z}
\end{eqnarray}
 where $M_{\rm DM}$ and $m_N$ denote the dark matter mass and nucleus mass,  respectively,
 $A_N$ and $Z_N$ are the mass number and proton number of the nucleus, 
 and $G_F$ is the Fermi constant~\cite{Arina:2007tm}.  
Here $\mu_X$ is the reduced mass defined by
 \dis{
\mu_X= \frac{\mdm m_X }{(\mdm+m_X)},
}
 and $m_n$ stands for the neutron mass.
In the expression (\ref{DD_Z}),
 $\sigma^Z_{\chi n}$ denotes the DM scattering cross section with a neutron,
 and we have used the fact that the $Z$ boson dominantly couples with a neutron (as opposed to a proton)
 as $(1-4\sin^2\theta_W)\simeq 0.076$, and hence we have
 neglected the contribution from scattering with a proton.

Usually the bound or signal of the direct detection experiments is given to
 the WIMP-nucleon scattering cross section, assuming the isospin-conserving case.
This is true for a conventional WIMP such as a neutralino, where Higgs boson-exchange processes are dominant.
For the $Z$-boson-mediated case, the DM interacts dominantly with a neutron,
 and thus the bound should be modified according to this.
Using \eq{DD_Z}, the corresponding WIMP-neutron cross section, $\sigma_n^{(Z)} $, for the $Z$-boson-mediated case is related to
 the isospin-conserving (IC) WIMP-nucleon scattering cross section, $\sigma_n^{\rm (IC)}$, by
\dis{
\sigma_n^{(Z)} =\sigma_n^{\rm (IC)}\bfrac{A}{A-Z}^2.
}
For Xenon $A\simeq130, Z=54$, and for Si in CDMS II $A=28, Z=14$. 
These factors give enhancement on the cross section by factors 4 and 3, respectively. 

\begin{figure}[!t]
  \begin{center}
  \begin{tabular}{c}
   \includegraphics[width=0.5\textwidth]{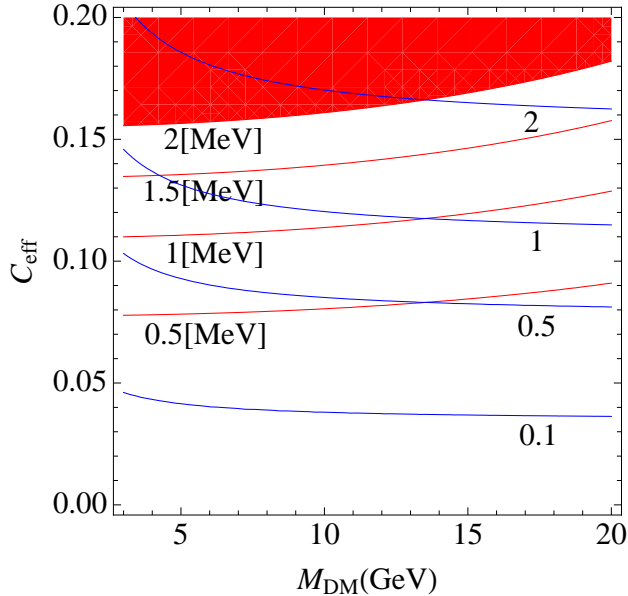}
     \end{tabular}
  \end{center}
  \caption{
The contours of the predicted scattering cross section with a neutron (blue)
  in $10^{-40} {\rm cm}^2$ and those of the extra $Z$-boson invisible decay width (red)
  as a function of sneutrino mass and $C_{\rm eff}$. The red region is disallowed by the LEP bound, $ \Delta \Gamma_{\rm inv}^Z< 2.0 \mev$~\cite{ALEPH:2005ab}.}
  \label{fig:ZinvAndDD}
\end{figure}

 In Fig.~\ref{fig:ZinvAndDD}, we show the contour of the $Z$-boson extra invisible decay width
 and the  WIMP-neutron scattering cross section in the plane of $C_{\rm eff}$ and the dark matter mass $M_{\rm DM}$.
 The contours of the predicted scattering cross section with a neutron (blue) are given 
  in units of $10^{-40} {\rm cm}^2$ with those of the extra $Z$-boson invisible decay width (red). The red region is disallowed by the LEP bound on the $Z$-boson extra invisible decay given in \eq{ZinvBound}.

\begin{figure}[!t]
  \begin{center}
  \begin{tabular}{c}
   \includegraphics[width=0.5\textwidth]{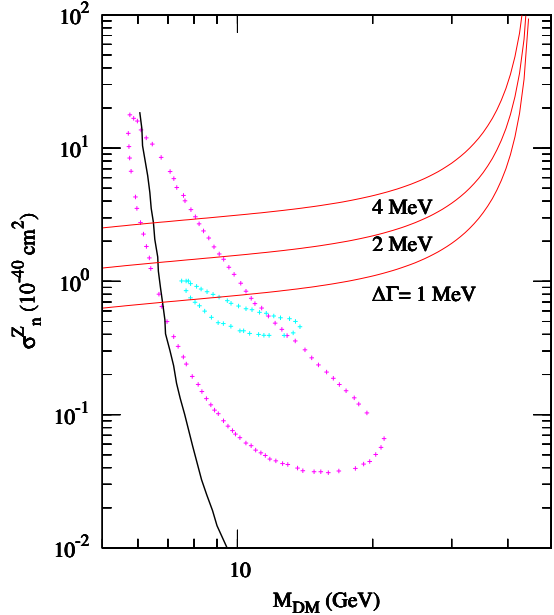}
     \end{tabular}
  \end{center}
  \caption{
The signal region and excluded region from direct dark matter searches [XENON100 (almost vertical line with black solid color), CDMS II(Si) (big closed loop of crosses with purple color), and CoGeNT (small closed loop of crosses with turquoise color)], and the magnitude of the corresponding Z-boson invisible decay width denoted by $\Delta \gamma = 1,2,4 $ MeV (red color).}
  \label{fig:Xsection}
\end{figure}

In Fig.~\ref{fig:Xsection}, we show the WIMP-neutron scattering cross section versus dark matter mass.
We show the constraint from XENON100~\cite{XENON100:2012},
 and the signals measured by CDMSII-Si~\cite{CDMSIISi} and CoGeNT~\cite{Kelso:2011gd}
 with the contour of the $Z$-boson extra invisible decay width.
Following Ref.~\cite{Frandsen:2013cna},
 we do not include the XENON10 limit in this paper to keep our discussion conservative.
We find that a still barely compatible region exists for a dark matter mass around $6\gev$ and
 the WIMP-nucleon cross section $\sigma_n^{(Z)} \simeq 10^{-40 } {\rm cm}^2$.

\section{Other constraints}
\label{other}

The discussion and conclusion in the previous section are model independent and
 were made applicable for any scalar DM scattering with a nucleon dominantly through $Z$-boson exchange by introducing the coefficient $C_{\rm eff}$.
In this section, we discuss other DM phenomenologies and experimental constraints.
To do this, we need to specify the particle model for Dirac sneutrino dark matter.

One model has been constructed
 with nonconventional supersymmetry (SUSY)-breaking mediation~\cite{ArkaniHamed:2000bq}.
Light sneutrino DM has been studied in Refs.~\cite{Belanger:2010cd,Dumont:2012ee}
 and has unfortunately turned out to be hardly compatible with LHC data,
 mainly due to the SM-like Higgs boson invisible decay width~\cite{Dumont:2012ee}.

There is another available model proposed by us~\cite{Choi:2012ap}
 in the context of the neutrinophilic Higgs doublet model~\cite{Ma,Wang,Ma:2006km,Nandi}.
Therefore in the rest of this section, as an example, we discuss
 other DM phenomenologies based on this model.

\subsection{Brief description of the model in Ref.~\cite{Choi:2012ap}}

The neutrinophilic Higgs model
 is based on the concept that the smallness of the neutrino mass might
 not come from a small Yukawa coupling
 but rather from a small vacuum expectation value (VEV) of the neutrinophilic Higgs field $H_{\nu}$.
As a result, neutrino Yukawa couplings can be as large as of the order of unity
 for a small enough VEV of $H_{\nu}$.
Other aspects-for instance, collider penomenology~\cite{Davidson:2009ha,Logan:2010ag,Haba:2011nb},
 astrophysical and cosmological consequences~\cite{Choi:2012ap,HabaSeto,Sher:2011mx,HSY},
 vacuum structure~\cite{Haba:2011fn},
 and variant models~\cite{HabaHirotsu,Haba:2012ai,Morita:2012kh}, -have also been studied.

The supersymmetric neutrinophilic Higgs model has 
 a pair of neutrinophilic Higgs doublets
 $H_{\nu}$ and $H_{\nu'}$ in addition to 
 up- and down-type two-Higgs doublets $H_u$ and $H_d$ in
 the MSSM~\cite{HabaSeto}.
A discrete $Z_2$ parity  is also introduced to discriminate
 $H_u (H_d)$ from $H_{\nu}(H_{\nu'})$,
 and the corresponding charges are assigned in Table~\ref{Table}. 
\begin{table}[t]
\centering
%\caption{$Z_{3}$-charge of each Field}
\begin{center}
\begin{tabular}{|c|c|c|} \hline
Fields  &  $Z_{2}$ parity & Lepton number \\ \hline\hline
MSSM Higgs doublets, $H_u, H_d$  &  $+$ &  0 \\ \hline
New Higgs doublets, $H_{\nu}, H_{\nu'}$ 
 &  $-$ & 0 \\ \hline
Right-handed neutrinos, $N$  &  $-$ & $1$ \\ \hline
\end{tabular}
\end{center}
\caption{The assignment of $Z_2$ parity and lepton number.}
\label{Table}
\end{table}
%
%\vspace{-5mm}
%\noindent
Under this discrete symmetry, 
 the superpotential is given by
\begin{eqnarray}
 W &=&y_{u} Q \cdot H_u U_{R}
 +y_d Q \cdot {H_d}D_{R}+ y_l L \cdot H_d E_{R} +y_{\nu} L \cdot H_{\nu} N  \nonumber \\
&& +\mu H_u \cdot H_d + \mu' H_\nu \cdot  H_{\nu'} 
+\rho H_u \cdot H_{\nu'} + \rho' H_\nu \cdot H_d ,
\label{superpotential}
\end{eqnarray}
 where we omit generation indices and dots represent the SU(2) antisymmetric product. 
The $Z_2$ parity plays a crucial role in 
 suppressing tree-level flavor-changing neutral currents and  
 is assumed to be softly broken 
 by tiny parameters of $\rho$ and $\rho' (\ll \mu, \mu' )$.
Here, we do not introduce lepton-number-violating Majorana mass for the RH neutrino $N$
 to realize a Dirac (s)neutrino.

By solving the stationary conditions for the Higgs fields, one finds that
 tiny soft $Z_2$-breaking parameters $\rho, \rho' $ generate  
 a large hierarchy of $v_{u,d} (\equiv \langle H_{u,d}\rangle) \gg 
 v_{\nu, \nu'}(\equiv \langle H_{\nu, \nu'}\rangle)$ expressed as
\begin{equation}
 v_{\nu} = {\cal O}\left(\frac{\rho}{\mu'}\right) v.
\end{equation}
It is easy to see that neutrino Yukawa couplings $y_{\nu}$ can be large
 for small $v_{\nu}$ using the relation of the Dirac neutrino mass $m_{\nu} = y_{\nu} v_{\nu}$.
For $v_{\nu} \sim 0.1 $ eV, it gives $y_{\nu} \sim 1$.
At the vacuum of $v_{\nu, \nu'}\ll v_{u,d}$,
 physical Higgs bosons originating from  $H_{u, d}$
 are almost decoupled from
 those from $H_{\nu,\nu'},$ except for a tiny mixing
 of the order of ${\cal O}\left(\rho/M_{\rm SUSY}, \rho'/M_{\rm SUSY} \right)$,
 where $M_{\rm SUSY} ( \sim 1 $ TeV) denotes the scale of soft SUSY-breaking parameters.
The former $H_{u,d}$ doublets almost constitute Higgs bosons in the MSSM --
 two $CP$-even Higgs bosons $h$ and $H$, 
 one $CP$-odd Higgs boson $A$, and a charged Higgs boson $H^\pm$ -- 
 while the latter, $H_{\nu, \nu'}$, constitutes 
 two $CP$-even Higgs bosons $H_{2,3}$,
 two $CP$-odd bosons $A_{2,3}$,
 and two charged Higgs bosons $H^\pm_{2,3}$. 
Thus, our model does not suffer from a large invisible decay
 width of SM-like an Higgs boson $h$ even for a large $y_{\nu}$ and a light lightest-supersymmetric-particle (LSP) dark matter.

At the vacuum, the mixing between left- and right-handed sneutrinos is estimated as
\begin{equation}
\sin \theta_{\tilde{\nu}} = {\cal O}\left( \frac{m_\nu}{M_{\rm SUSY}} \right) .
\label{LRmixing}
\end{equation}
We find that the RH sneutrino $\tilde{N}$ has very suppressed interactions
 with the SM-like Higgs boson or $Z$ boson at tree level,
 since they are proportional to the mixing of left-handed and RH neutrinos'
 $\sin \theta_{\tilde{\nu}} $ in \eq{LRmixing}.
However, radiative corrections induce a sizable coupling between RH sneutrinos and the $Z$ boson.
We have parametrized the effective interaction between the RH sneutrino DM and $Z$ boson by $C_{\rm eff}$;
 then, the vertex induced by the scalar ($H_{\nu}$-like Higgs boson and $\tilde{\nu}_L$)
 loop~\footnote{The fermion loop contribution is suppressed by helicity.} is given as
\begin{eqnarray}
{\rm Vertex} &=& \frac{g}{2\cos\theta_W}(k_1^\mu+ k_2^\mu) C_{\rm eff},
\end{eqnarray}
with
\begin{eqnarray}
C_{\rm eff} &=& \frac{(-i)(y_\nu A_\nu)^2}{12(4\pi)^2 M^2},  
\label{Ceff}
\end{eqnarray}
 where $k_1^\mu$ and $k_2^\mu$ are the ingoing and outgoing momenta of the RH sneutrino
 and for simplicity we take equal masses for particles in the loop, $M = M_{H_\nu}= M_{\tilde{\nu}_L}$.

%
%Therefore with only this single diagram,
%\dis{
% |C_{eff}|^2 = 2.8\times 10^{-7} \left[ \frac{(y_\nu A_\nu)^2}{M^2}  + ... \right]^2,
%}
% 

By comparing Fig.~\ref{fig:Xsection} and~\eq{DD_Z} with Eq.~(\ref{Ceff}), 
%
%we find that
%\dis{
%\sigma^Z_{\chi n} \simeq |C_{\rm eff}|^2 \frac{G_F^2}{2\pi} \simeq 10^{-40 } {\rm cm}^2=2.6\times 10^{-13} \gev^{-2}.
%}
%Therefore
 we find the parameter set 
\dis{
y_\nu \,A_\nu \simeq 14.4 \, M \qquad {\rm and} \qquad M_{\rm DM} \simeq 6 \,{\rm GeV},
\label{set}
}
 can explain the CDMS II Si result.

\subsection{Annihilation cross section}

\begin{figure}[!t]
  \begin{center}
  \begin{tabular}{c}
   \includegraphics[width=0.5\textwidth]{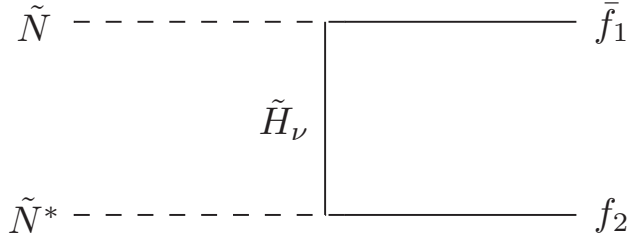}
     \end{tabular}
  \end{center}
  \caption{Tree-level diagram for the annihilation of RH sneutrinos.}
  \label{fig:Tree}
\end{figure}

The dominant tree-level annihilation mode of $\tilde{N}$ in the early Universe 
 is the annihilation into a lepton pair
 $\tilde{N} \tilde{N}^*\rightarrow \bar{f_1} f_2 $ mediated by the heavy $H_{\nu}$-like
 Higgsinos as described in Fig.~\ref{fig:Tree}. 
The final states $f_1$ and $f_2$ are charged leptons for the $t$-channel $\tilde{H}_{\nu}$-like
 charged Higgsino ($\tilde{H}_{\nu}^\pm$) exchange, while
 thy are neutrinos for the $t$-channel $\tilde{H}_{\nu}$-like
 neutral Higgsino ($\tilde{H}_{\nu}^0$) exchange. 
The thermal averaged annihilation cross section for this mode
 in the early Universe when using the partial wave expansion method is given by~\cite{Lindner:2010rr}
\begin{equation}
\langle \sigma v\rangle_{f\bar{f}} = \sum_f \left(
 \frac{y_\nu^4}{16\pi} \frac{m_f^2}{(M^2_{\tilde N} + M^2_{\tilde{H}_\nu} )^2}
 + \frac{ y_{\nu}^4}{8 \pi }
    \frac{M^2_{\tilde{N}}}{ ( M_{\tilde{N}}^2 + M_{\tilde{H_{\nu}}}^2 )^2 }\frac{T}{ M_{\tilde{N}} } + ... \right) ,
    \label{treelevel}
\end{equation}
 where we used $\VEV{v_{\rm rel}^2}=6T/M_{\rm DM}$
 with $v_{\rm rel}$ being the relative velocity of annihilating dark matter particles,
 $m_f$ is the mass of
 the fermion $f$, and $M_{{\tilde H}_\nu} \simeq \mu'$
 denotes the mass of the $\tilde{H}_{\nu}$-like Higgsino.
For simplicity we have assumed that Yukawa couplings are universal for each flavor.
Since the $s$-wave contribution of the first term on the right-hand side is helicity suppressed,
 the $p$-wave annihilation cross section of the second term is relevant for the dark matter relic density
 at the freeze-out epoch.

In the neutrinophilic Higgs model, the sneutrino has -- in addition to the
 tree-level processes -- a sizable annihilation cross section into two photons through a one-loop diagram, which has been pointed out in Ref.~\cite{Choi:2012ap}.
The charged components of the $H_\nu$ scalar doublet and charged scalar fermions make the triangle or box 
 loop diagram, and the two photons can be emitted from the internal charged particles.
%In this case, the photons have line spectrum with energy 
%%
%\dis{
%E_{2\gamma} = M_{\tilde N}.
%}
%%
%with the amplitude 
%\dis{
%{\cal M}_{2\gamma} = {\cal M}^{\text{Tringle}}_{2\gamma}+{\cal M}^{\text{Box}}_{2\gamma}.
%\label{twogamma}
%}
%Since DM is nonrelativistic we can ignore the momentum of DM. 
For the mass spectrum we are interested in now, $M_{H_{\nu}}, M_{\tilde l}\gg M_{\tilde N}$,
 we obtain the annihilation cross section to two photons via one loop as
\begin{eqnarray}
\VEV{\sigma v}_{2\gamma}
%&=&\frac{|M|^2_{2\gamma} }{32\pi M^2_{\tilde N} }
 &\simeq& \frac{ \alpha^2_{\rm em}}{8\pi^3}\frac{y_\nu^4 (A_\nu^2+\mu'^2)^2}{M_{\rm ch}^4} \frac{4}{M^2_{\tilde N}} \nonumber \\
 &=& 2.8\times10^{-8} \, \gev^{-2}  \bfrac{6\gev }{M_{\tilde N}}^2 \frac{y_\nu^4 (A_\nu^2+\mu'^2)^2}{M_{\rm ch}^4} ,
\end{eqnarray}
 where we have used  $M_{H_\nu}=M_{H_\nu'}=M_{\tilde l}\equiv M_{\rm ch}$ for simplicity.

Therefore for the total annihilation cross section of RH sneutrino DM, we obtain 
\begin{equation}
\VEV{\sigma v} = \VEV{\sigma v}_{f\bar{f}} + \VEV{\sigma v}_{2\gamma} .
\end{equation}

Now if we attempt to reproduce the latest CDMS II-Si data by taking a parameter
 set given by Eq.~(\ref{set}),
 we find that two-photon production via one loop is dominant and thus
\dis{
\VEV{\sigma v} \simeq \VEV{\sigma v}_{2\gamma} \simeq 10^{-3} \, \gev^{-2},
\label{AnnihXsection} 
}
for the given parameters in \eq{set}.
This loop-induced annihilation does not only dominate the tree-level annihilation
 but also exceeds the standard value $\VEV{\sigma v} \simeq 10^{-9} \, \gev^{-2}$.
This DM appears to not have the correct thermal relic abundance if the relic density is determined
from its thermal freeze-out.

\subsection{Dark matter relic abundance and indirect DM search constraints}

As stated above, from Eq.~(\ref{AnnihXsection}) we see that 
 the standard thermal relic density of $\tilde{N}$ with zero chemical potential 
 leads to a too small value for $\Omega h^2 \ll 0.1$.
However, we know that our Universe is baryon asymmetric.
Hence, we expect that lepton asymmetry is also nonvanishing.
In fact,
 the sphaleron process, which interchanges baryons and leptons, plays an important role
 in many baryogenesis mechanism
 and leaves a similar amount of baryon asymmetry and lepton asymmetry.

Because our model is supersymmetric, a promising mechanism would be
 Affelck-Dine (AD) baryo(lepto)genesis~\cite{AffleckDine}.
Candidates for a promissing AD field $\phi$
 are, e.g., $\bar{u}\bar{d}\bar{d}$ or $LL\bar{e}$ directions with the
 nonrenormalizable superpotential $\Delta W = \phi^6/M^3$,
 where $M$ is a high cutoff scale for this operator.
The generated baryon $(q=B)$ or lepton $(q=L)$ asymmetry for those directions 
 have been studied by many authors and evaluated as~\cite{ADpapers,EMD,FH,Seto,RS,SY}
\begin{equation}
\frac{n_q}{s} \simeq 10^{-10} q \sin\delta\left(\frac{A_{\phi}}{1 \rm{TeV}}\right)
 \left(\frac{1 \rm{TeV}}{m_{\phi}}\right)^{3/2}\left(\frac{T_R}{10 \rm{TeV}}\right)
 \left(\frac{M}{10^{-2} M_P}\right)^{3/2},
\end{equation}
 for a relatively low reheating temperature after inflation $T_R$
 in gravity-mediated SUSY-breaking models,
 where $m_\phi$ and $A_{\phi}$ are soft SUSY-breaking mass and A term for the AD field, 
 $\delta$ is an effective $CP$ phase, $M_P$ is the reduced Planck mass, and
 $M$ is taken to be around the grand unification scale.
Then, the charge of $Q$-balls, even if they are formed, is small enough
 for a $Q$-ball to evapolate quickly~\cite{Banerjee:2000mb} and to not affect the dark matter
 density.~\footnote{If unstable $Q$-balls survive the evapolation,
 the decay generates LSP dark matter, as discussed in Ref.~\cite{EMD,FH} for neutalino
 and Refs.~\cite{RS,SY} for axino.}
To be precise, this generated $B-L$ asymmetry is related
 to the baryon and lepton asymmetry generated by the sphaleron process~\cite{Kuzmin:1985mm},
\begin{eqnarray}
\frac{n_B}{s} \sim \frac{n_L}{s} = {\mathcal O}(10^{-10}) .
\end{eqnarray}
Since a Dirac sneutrino carries a lepton number and has a large annihilation cross section
 [as in Eq.~(\ref{AnnihXsection})],
 our sneutrino is one of the natural realizations of the so-called asymmmetric dark matter
 (ADM)~\cite{Barr:1990ca,Barr:1991qn,Kaplan:1991ah,Thomas:1995ze,Hooper:2004dc,Kitano:2004sv,Kaplan:2009ag}, and in our model
 only $\tilde N$ remains after annihilation with $\tilde N^*$. 
Thus, the relic abundance is actually determined by its asymmetry and the mass.

For a novanishing sneutrino asymmetry similar to the baryon asymmetry,
\dis{
Y_{\tilde N}\equiv \frac{n_{\tilde N} -n_{\tilde N^*}}{s} = {\cal O}( 10^{-10}),
} 
 and a mass of about $5-6$ GeV, the correct relic density for dark matter
 $\Omega_{\tilde N}h^2\simeq 0.1$ is obtained.

%The Planck satellite gives the relic density of cold dark matter and baryon as
%\dis{
%\frac{\Omega_{CDM}}{\Omega_{Barion}}=5.45.
%}
%Considering the mass of proton and dark matter
%the ratio of the number density at present is
%\dis{
%Y_{CDM}=5.4\times \frac{m_B}{m_{DM}}Y_{Baryon}\simeq 0.9 Y_{Baryon}
%}

Finally, we note that
 our model is free from any indirect search for DM annihilation; in other words,
 DM annihilation cannot produce any signal
 because of the ADM property, namely, the absence of anti-DM particles in our Universe.

\section{Conclusion}
\label{conclusion}

We have shown that
 mostly right-handed Dirac sneutrinos are a viable supersymmetric light
 DM candidate and
 have a large enough cross section with nucleons to account for 
 possible signals observed at direct DM searches.
The $Z$-boson-mediated scattering does not relax the tension 
 among direct DM search experiments 
 and
 is constrained by the invisible decay width of the $Z$ boson.
Nevetherless, we have found that
 a part of the signal region for CDMS II(Si) events
 remains outside the excluded region by XENON100.
As an example of specific particle models,
 we have shown that a Dirac right-handed sneutrino
 with neutrinophilic Higgs doublet fields is a viable light dark matter candidate.

\section*{Acknowledgments}

K.-Y.C. was supported by the Basic Science Research Program through the National Research Foundation of Korea (NRF) funded by the Ministry of Education, Science and Technology Grant No. 2011-0011083.
K.-Y.C. acknowledges the Max Planck Society (MPG), the Korea Ministry of
Education, Science and Technology (MEST), Gyeongsangbuk-Do and Pohang
City for the support of the Independent Junior Research Group at the Asia Pacific
Center for Theoretical Physics (APCTP).

%======================================%
%<<<<<<<<<<< appendix >>>>>>>>>>>>>%
%======================================%
%======================================%
%<<<<<<<<<<< bibliography >>>>>>>>>>>>>%
%======================================%

%%%%%%%%%%%%%%%%%%%%%%%%%%%%%%%%%%%%%%%%%%%%%%%%%%%%%%%%%%%%

\end{document}